\documentclass[superscriptaddress,twocolumn,amsmath,amssymb,aps,pra,showpacs,showkeys]{revtex4-1}

\usepackage{graphicx}
\usepackage{dcolumn}
\usepackage{bm}
\usepackage{color}
\usepackage{tabularx}
\usepackage{epsfig}
\usepackage{amsmath}
\usepackage{amssymb}
\usepackage{graphicx}
\usepackage{wasysym}
\usepackage{dcolumn}
\usepackage{bm}
\usepackage{subfigure}
\usepackage{psfrag}
\usepackage{subfigure}
\usepackage{times}

\definecolor{yblue}{rgb}{0.06, 0.3, 0.57}
\usepackage[pdftex]{hyperref}
\hypersetup{colorlinks=true,linkcolor=yblue,citecolor=yblue,urlcolor=yblue}

\begin{document}

\title{How small-world interactions can lead to improved quantum annealer designs}

\author{Helmut G.~Katzgraber}
\affiliation{Microsoft Quantum, Microsoft, Redmond, WA 98052, USA}
\affiliation{Department of Physics and Astronomy, Texas A\&M University,
College Station, Texas 77843-4242, USA}
\affiliation{1QB Information Technologies (1QBit), Vancouver, British
Columbia, Canada V6B 4W4}
\affiliation{Santa Fe Institute, 1399 Hyde Park Road, Santa Fe, New Mexico 
87501, USA}

\author{M.~A.~Novotny}
\email{man40@msstate.edu}
\affiliation{Department of Physics and Astronomy, Mississippi State
University, Mississippi State, MS 39762-5167, USA}
\affiliation{HPC$^2$ Center for Computational Sciences, Mississippi
State University, Mississippi State, MS 39762-5167, USA}
\affiliation{Faculty of Mathematics and Physics, Charles University, 
Ke Karlovu 5, CZ-121 16 Praha 2, Czech Republic}

\date{\today}

\begin{abstract}

There are many factors that influence the design of quantum annealing
processing units. Here we address the issue of improving quantum
annealing processing unit designs from the point of view of the critical
behavior of spin glasses. It has been argued [Phys.~Rev.~X {\bf 4},
021008 (2014)] that among the most difficult Ising spin-glass
ground-state problems are those related to lattices which exhibit a
finite-temperature spin-glass transition.  Here, we show that adding
small-world couplers between qubits (spins) to the native quasi-planar
quantum processing unit graph results in a topology where a disordered
Ising system can undergo a finite-temperature spin-glass transition,
even when an Ising spin glass on the quasi-planar native graph does not
display a transition into a glassy phase at any finite temperature. To
ensure that these systems can be engineered with current fabrication
techniques, using large-scale Monte Carlo simulations we demonstrate
that highly-constrained systems restricted to a few fabrication layers
and with fixed coupler angles can also exhibit a finite-temperature
spin-glass transition. This indicates that these systems might be
mean-field-like, which also means that embedding highly-nonplanar
problems might be simplified when compared to the underlying native
topology. Our results are illustrated using the quasi-planar Chimera
topology currently used in the D-Wave Systems Inc.~quantum annealing
machines, as well as standard two-dimensional square lattices. The
presented approach can be generalized to other topologies.

\end{abstract}



\maketitle

\section{Introduction}

We are currently in the midst of an exciting and formative era in the
emerging field of quantum computing.  The first quantum computers are
being built and rapidly refined.  Quantum computing is rapidly
transitioning from a theoretical curiosity to being a practical,
indispensable tool on the computational science, engineering, and
business workbench.  A quantum computer operates on qubits (quantum
bits) via quantum operations, as opposed to a classical computer which
operates on bits (the binary values $0$ and $1$).  A register of $N$
qubits may simultaneously be in a quantum superposition state of all
$2^N$ possible states \cite{mermin:07}.  In contrast, a classical
computer must always be in one of the possible $2^N$ states.  The
ability to operate on qubits rather than bits is what gives the enticing
game-changing potential possibilities to quantum computers. Quantum
computers hold the promise to be able to have an exponential increase in
the ability to calculate solutions to certain problems, as compared with
classical computers \cite{mermin:07,devries:12,rieffel:14}.

In any emerging technology, the architectures and underlying materials,
as well as the actual engineering process, are developed and improved
alongside the first operational machines. For example, in 2016 IBM
publicly released a gate-model quantum computing device with five qubits
on a star-shaped topology, while, in parallel, an internal version with
$16$ qubits and a different topology (released in 2017) was being tested
\cite{corcoles:15}. Similarly, whenever D-Wave Systems Inc.~releases a
new quantum processing unit for its quantum annealing machines,
different hardware iterations and graph topologies are tested with 
different metrics that, likely, range from problem embedability, qubit
cross talk, engineering constraints, and coherence times, to
application-driven metrics, to name a few.

While digital quantum processing units can be programmed to study a
variety of problems, a quantum annealer's main purpose is to find optima
for hard binary optimization problems via quantum annealing
\cite{kadowaki:98,kadowaki:98a,farhi:00,farhi:01,morita:08,das:08,albash:16a}.
The architecture of the quantum annealing machine may be viewed as the
arrangement of qubits at the vertices (nodes) of a graph, and the bonds
of the undirected graph are the couplers (or interconnects) between the
qubits. Programming a quantum annealer consists in providing a bias
field $h_i$ for each qubit $i$, as well as the magnitude and sign of the
interactions between pairs of qubits $i$ and $j$, the coupling strength
and sign of the coupler $J_{ij}$. Any problem in a quadratic binary
unconstrained format can then be embedded onto the device's topology.
For native problems (i.e., without embedding) the nonplanarity of the
quantum annealer graph, together with both possible signs of the
coupling between qubits, can lead to frustration in the encoded
spin-glass-like problem, which can potentially make the finding of the
ground state an NP-hard problem \cite{barahona:82,barahona:82b}.

Recently, multiple types of synthetic benchmark problems have emerged
\cite{hen:15a,king:15,denchev:16,mandra:16,mandra:17a} in an effort to
demonstrate that D-Wave quantum annealing machines can outperform
algorithms on classical CMOS hardware. Although to date only a
verifiable constant speedup has been found \cite{mandra:17b} and
applications have yielded mixed results
\cite{perdomo:12,perdomo:15b,rieffel:15,venturelli:15b,mandra:16c,perdomo:17y},
much effort is still be expended to elucidate the application scope of
the D-Wave device. While synthetic benchmarks designed to ``break''
classical algorithms are excellent benchmark problems, their
most-prominent drawback is the need to encode a hard logical problem
into the physical hardware layout of the quantum annealing processing
unit (QAPU). This overhead often results in considerably smaller logical
problems that might be far from the asymptotic regime where the true
scaling sets in. As such, it is desirable to develop native benchmark
problems that use all qubits on the QAPU and that are computationally
tunable and ideally hard. While this is not easily accomplished with
current hardware \cite{katzgraber:14}, it might be possible with better
hardware designs, as proposed in this work.

Defining and detecting quantum speedup is an intricate task
\cite{ronnow:14}. Quantum speedup is one component in the search for
quantum supremacy. Quantum supremacy may be defined as the point where a
quantum computer can perform tasks more efficiently than a classical
computer. This can be, for example, faster than current computer
technologies, using considerably less memory, or simply tasks that
cannot be simulated on classical hardware at all. As the problem size
and classical computer hardware are scaled up, it is expected that the
resource requirements for quantum hardware will be considerably less
(i.e., the quantum computer outperforms classical hardware by an even
larger margin). For quantum annealers, being able to more efficiently
embed problems (i.e., with a smaller embedding overhead) that might show
either any quantum advantage or speedup in the native architecture of
the QAPU plays a central role in the assessment of the potential of such
devices. It is this component on which we concentrate in this article.

\begin{center}
\begin{figure}[tb]
\includegraphics[width=0.30\textwidth]{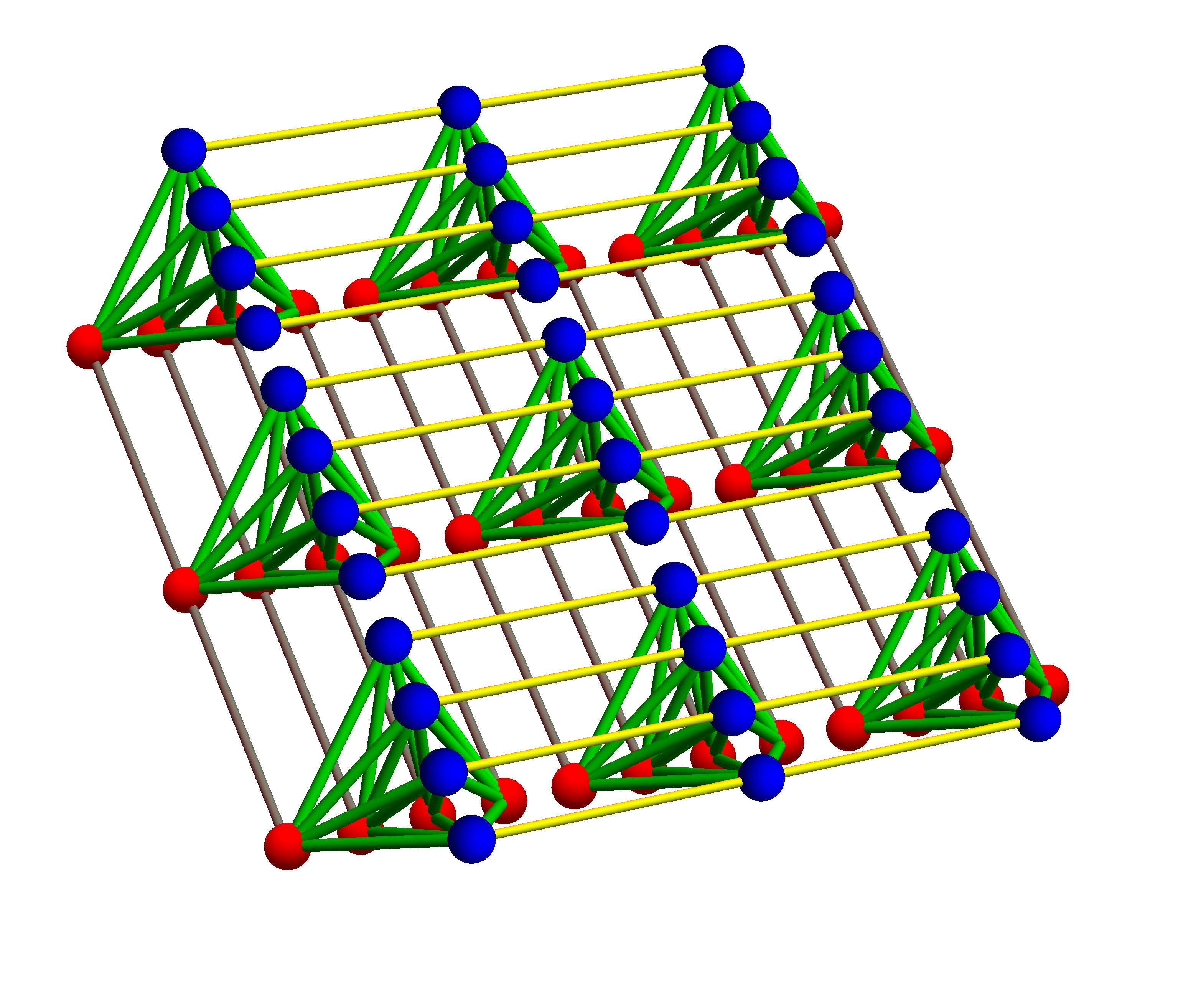}
\\
~~~
\\
\includegraphics[width=0.95\columnwidth]{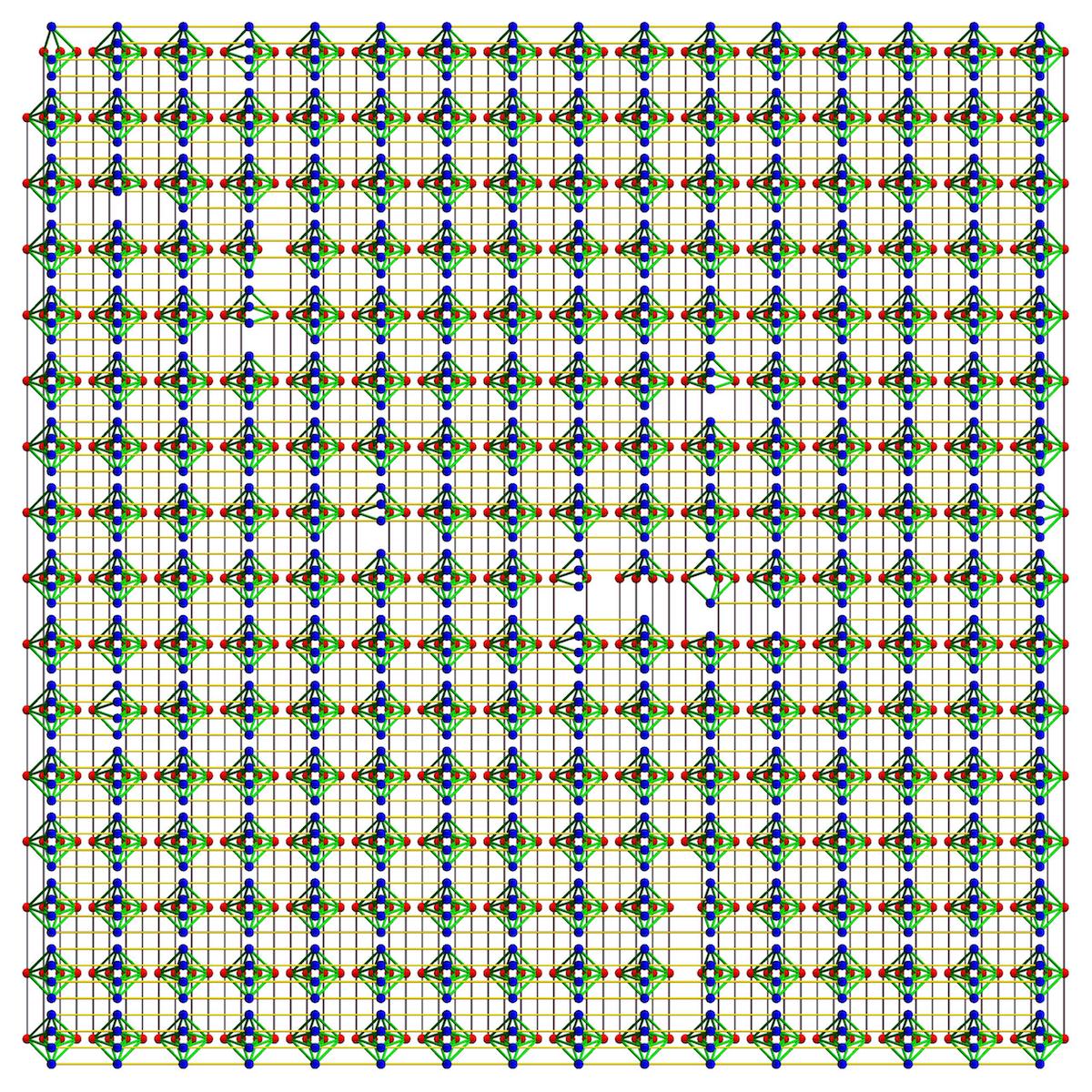}
\caption{
\label{Fig:SW:Fig01}
$K_{4,4}$ Chimera lattices with $N=8 L^2$ sites, illustrated as a
quasi-two-dimensional (referred to as ``two-level'') system.  Every unit
cell has four qubits in the upper layer (blue spheres) and four qubits
in the lower layer (red spheres), and the intra-unit-cell couplers have
each of the top four qubits connected to every one of the four lower
sites (green cylinders).  The inter-unit-cell connections in the top
layer (yellow cylinders) and the bottom layer (gray cylinders) are
shown. The graph has free boundary conditions. The top panel depicts a
Chimera graph with $L=3$ and the bottom panel depicts the Chimera graph
of a current 2000Q model (top view) with $L=16$ and some missing qubits
and couplers (of the $N = 2048$ available qubits only $2023$ are
operational). The Supplemental Material contains three-dimensional
interactive versions of these graphs \cite{comment:supp}.
}
\end{figure}
\end{center}

Although there are many metrics that, in principle, drive the design of
a particular quantum chip topology, because no definite
application-based signs of quantum speedup exist to date, we focus on
well-understood {\em native} \cite{comment:native} fundamental
spin-glass benchmarks in an effort to ``tickle'' any advantage out of
the available hardware. It has been argued 
\cite{katzgraber:14} that Ising spin-glass problems might pose greater
challenges to optimization machines, as well as classical algorithms, if
the underlying optimization problem undergoes a finite-temperature
spin-glass transition where the barriers in the free-energy landscape
are more prominent at low close-to-zero temperatures. Similar arguments
hold for graphs with higher connectivity \cite{venturelli:15a}. Quantum
processing units tend to have a planar topology, mainly due to
fabrication constraints. Because spin glasses on quasi-two-dimensional
topologies---such as the $K_{4,4}$-based Chimera graph \cite{bunyk:14}
of the D-Wave devices---do not have a finite-temperature spin-glass
transition, here we outline an approach to induce finite-temperature
transitions, as well as mean-field-like behavior for spin glasses on
quasiplanar topologies by adding (constrained) small-world (SW) couplers
\cite{watts:98,hastings:03}. While a finite-temperature spin-glass
transition allows for the development of more-elaborate benchmarking
techniques, the small-world properties of the QAPU topology that lead to
mean-field critical behavior potentially assist with the embedding of
nonplanar problems in the hard-wired topologies of these systems. We
illustrate the proposed approach using D-Wave's Chimera graph (see
Fig.~\ref{Fig:SW:Fig01}) and demonstrate that the approach is generic by
also studying a two-dimensional square lattice.

\begin{center}
\begin{figure}[tb]
\vspace{0.05 cm}
\includegraphics[width=0.90\columnwidth]{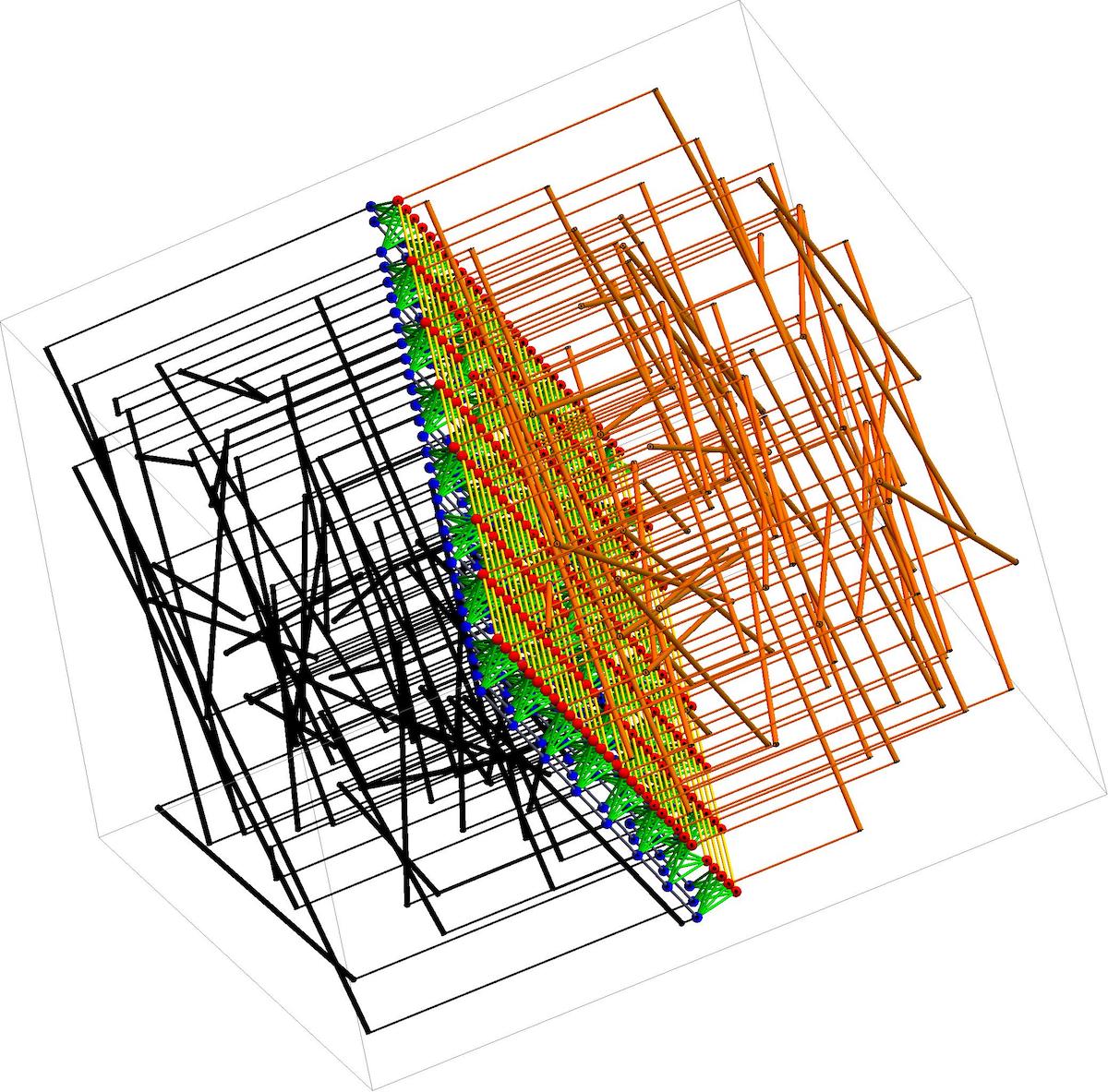}
\caption{
\label{Fig:SW:Fig02}
A $K_{4,4}$ Chimera lattice with $L=8$ ($N = 512$ sites) and with added
small-world couplers, illustrated as a two-level system.  Rendered is an
actual D-Wave Two chip with $489$ working qubits and $1345$ operational
couplers.  Every unit cell has four qubits in the upper layer (blue
spheres) and four qubits in the lower layer (red spheres), together with
each of the top four qubits connected to every one of the four lower
qubits (green cylinders). The inter-unit-cell connections in the top
layer (yellow cylinders) and the bottom layer (gray cylinders) are
shown. The graph has free boundary conditions.  Qubits in the top
(bottom) layer have an additional $64$ small-world couplers depicted as
orange (black) cylinders. The only constraint on the small-world
couplers is that each qubit can have at most one additional small-world
coupler.  To keep the connections from overlapping, every small-world
coupler is at a different height above or below the qubit planes. The
Supplemental Material contains a three-dimensional interactive version of
this graph \cite{comment:supp}.
}
\end{figure}
\end{center}

We argue that the graph of any QAPU should allow one to study
optimization problems that map onto a spin-glass model above the upper
critical dimension $d_{\rm upper}$. For an Ising spin glass, $d_{\rm
upper}=6$ \cite{binder:86}. For a graph with $d\ge d_{\rm upper}$ the
critical exponents are those of a mean-field model; namely, the same as
having every qubit interacting with every other qubit \cite{binder:86}.
A graph that has a three-dimensional cubic structure, for example, will
never be able to embed a computational problem which corresponds to a
six-dimensional hypercube without a large overhead.  Unfortunately, for
mostly engineering reasons, building a QAPU with a hypercube in six or
more dimensions is not feasible at the moment.  The approach presented
here whereby SW connections \cite{watts:98} are added to the underlying
problem graph potentially enables an economic route to build QAPUs for
problems that map onto spin glasses on topologies that are above the
upper critical dimension.  This small-world-enhanced architecture is
related to the architectures of perfectly scalable classical massively
parallel computers \cite{korniss:03,comment:patent-sw}. Our study thus
enables us to establish a reasonable requirement for the architecture of
any QAPU \cite{comment:patent-aqc}.  Trying to engineer a QAPU with
unconstrained small-world interactions may well be unrealistic (see
Fig.~\ref{Fig:SW:Fig02}) because for $k$ randomly placed additional
small-world couplers worst-case $k$ additional fabrication layers are
needed.  Consequently, a large portion of this paper is devoted to
investigating whether constrained small-world connections are sufficient
to obtain both a finite-temperature spin-glass transition and the
associated mean-field spin-glass critical exponents.

For ferromagnetic models, the addition of small-world bonds has been
studied both theoretically and by computer simulations
\cite{scalettar:91a,gitterman:00,barrat:00,pekalski:01,hong:02,herrero:02}.
The small-world bonds can be added by starting, for example, with either
a one-dimensional or a two-dimensional regular lattice.  The consensus
that has emerged is the addition of the small-world bonds gives a
finite-temperature transition, even when one starts from a regular
one-dimensional lattice that does not have a finite critical
temperature, and furthermore the exponents associated with the critical
behavior \cite{hastings:03,zhang:05,zhang:06} have mean-field behavior;
namely, the same as that associated with ferromagnetic models above the
upper critical dimension $d_{\rm upper}^{{\rm ferro}}=4$.  For the
ferromagnet, finite-size scaling of both the ferromagnetic Binder
cumulant and ferromagnetic susceptibility gives the mean-field exponents
$\nu_{d\ge4}^{{\rm ferro}}=2$ and $\gamma_{d\ge 4}^{{\rm ferro}}=1$.
More information on additional features of small-world networks can be
found in
Refs.~\cite{watts:98,watts:99,albert:99,albert:02,barabasi:16,newman:18}.

This paper is structured as follows. Section \ref{sec:model} introduces
the Ising spin-glass Hamiltonian, the graphs used in our studies
(Chimera and square lattices with and without small-world connections),
and the computational methods used to study a potential spin-glass
transition using finite-size scaling.  Section \ref{sec:SWno} contains
results for unconstrained small-world connections added to Chimera
graphs. Results of adding small-world connections to Chimera graphs with
constraints are presented in Sec.~\ref{sec:SWlayer} for constraints to a
small number of layers and in Sec.~\ref{sec:SWlayerangle} for
constraints also confined to angles of $\pm\pi/4$.  Section
\ref{sec:SWsquare} contains results for a square lattice with added
connections constrained to both a small number of layers and to angles
of $\pm\pi/4$. A discussion and conclusions are presented in
Sec.~\ref{sec:DisCon}.

\section{\label{sec:model} Model and Methods}

We consider a graph $G$ with $N$ nodes, where the QAPU qubits reside on
the vertices of the graph. Graph edges represent the couplers between
the qubits. We study a spin glass on the native topology of the QAPU;
that is,
\begin{equation}
\label{Eq:Ham}
{\cal H} = - \sum\limits_{i,j=1}^N J_{ij} S_i S_j ,
\end{equation}
where, $S_j \in \{\pm 1\}$ represent Ising spins. The couplers $J_{ij}$
of the native graph are chosen from a Gaussian distribution with zero
mean and standard deviation of $1$. The additional small-world couplers
(explained in detail below) are chosen from a bimodal distribution
(i.e.,  $J_{ij} \in \{\pm 1\}$ with equal probability to somewhat
accommodate the fact that the engineering and calibration of the
additional small-world couplers is more complex than that of the native
couplers of the native QAPU graph).  We work only with $2$-local graphs
in this paper, leaving a discussion of $k$-local ($k > 2$) graphs to
Sec.~\ref{sec:DisCon}.  The (nonplanar) graphs studied have cycles
(closed paths along the bonds) and the couplings can have both signs; in
general, therefore, not all bonds can simultaneously be satisfied by an
arrangement of the $N$ values of $S_j$, thus leading to frustration.
This also means that finding the ground-state of these systems is an
NP-hard problem \cite{barahona:82,barahona:82b}.  Because the existence
of a stable spin-glass state in a field for short-range systems is
controversial \cite{young:04,katzgraber:05c,joerg:08a,katzgraber:09c},
without loss of generality we focus here on spin glasses without local
biases (fields).

We note, however, that nonzero biases (which, incidentally, could lead
to interesting benchmark designs in the search for quantum speedup) play
an important role in applications. In Ising systems, the introduction of
biases can change the complexity class of the problem.  Mean-field Ising
spin glasses, however, exhibit a finite-temperature spin-glass
transition for biases and temperatures below the de Almeida-Thouless
line \cite{almeida:78}. This means that for small-enough biases, a
spin-glass phase will be present when the system is in the mean-field
regime (i.e., when small-world couplers are added).  While infinitesimal
biases would destroy the fragile spin-glass state on short-range
topologies such as the Chimera topology, the addition of small-world
couplers would make the resulting spin-glass state on such lattices
stabler against local qubit noise (i.e., random biases).  Similarly, if
the critical scaling is mean-field-like for zero-bias fields, one
expects the same for the case with bias fields, as both have the same
spin-glass upper critical dimension $d_{\rm upper}=6$.

For a spin glass, the canonical order parameter is the spin overlap
defined as 
\begin{equation} 
\label{Eq:Osg} 
q = 
\frac{1}{N} \sum\limits_{j=1}^N S_j^{(\alpha)} S_j^{(\beta)}  ,
\end{equation} 
where, $(\alpha)$ and $(\beta)$ represent two copies of the
system with the same disorder.  It is convenient to construct
dimensionless quantities from the order parameter in Eq.~\eqref{Eq:Osg}
to better pinpoint the location of a phase transition. The Binder ratio
is such a dimensionless function, which means that, at a putative
transition, data for different system sizes $N$ will cross when $T =
T_c$, where $T_c$ is the critical temperature (up to
corrections to scaling).  Defined via
\begin{equation}
\label{Eq:Binder}
g \> = \> \frac{1}{2} \> 
\left(3-\frac{[\langle q^4\rangle]_{\rm av}}{[\langle q^2\rangle]_{\rm av}^2}\right) \>,
\end{equation}
the Binder ratio scales as
\begin{equation}
\label{Eq:gScaleff}
g \sim G\left[ N^{1/\nu_{\rm eff}} \left(T-T_c\right)\right]
\equiv G\left( x \right),
\end{equation}
where $x = N^{1/\nu_{\rm eff}} \left(T-T_c\right)$.  In
Eq.~\eqref{Eq:Binder} the angular brackets $\langle \cdots \rangle$
represent a thermal average and the square brackets $[\cdots]_{\rm av}$
a configurational average. The configurational average is over
coupler values when only the native graph is studied.  However, in the
case where small-world couplers without restrictions are added to the
native graph, the average is over coupler values, as well as the
position of the small-world couplers.  For the restricted
simulations a {\em fixed} set of small-world couplers is used (i.e.,
again the configurational average is over only the values of the
couplers).  The effective critical exponent $\nu_{\rm eff}$ is given by 
\begin{equation}
\label{Eq:neEff}
\nu_{\rm eff} = 
\left\{
\begin{array}{lcl}
\nu & \qquad\qquad & d_{\rm lower}<d\le d_{\rm upper} \\
\nu_{\rm MF} & & d\ge d_{\rm upper} . \\
\end{array}
\right .
\end{equation}
In Eq.~\eqref{Eq:neEff} $\nu_{\rm MF} = 3$, where MF stands for ``mean
field.'' For spin glasses, the upper critical dimension is $d_{\rm
upper} = 6$ \cite{binder:86,parisi:97a} and the lower critical dimension
$d_{\rm lower}$ lies between $2$ and $3$
\cite{boettcher:05d,demirtas:15}, with a commonly used value being $5/2$
\cite{boettcher:05d}.  From Eq.~\eqref{Eq:gScaleff} the critical
parameters $\nu_{\rm eff}$ and $T_c$ can be estimated via a
finite-size scaling analysis \cite{melchert-autoscale:09}.  For a
hypercubic lattice with short-range interactions $L=N^{1/d}$, with $d$
the space dimension. Often finite-size scaling equations are written in
terms of $L$.  Here, however, because of the added small-world couplers
a space dimension $d$ is somewhat ill defined and thus we write all
expressions in terms of $N$.

In a mean-field-like system the critical exponent $\eta$ associated with
the susceptibility is known {\em a priori} and can be expressed in terms of
$\nu_{\rm MF}$. Therefore, to obtain an independent estimate of a
transition we also study the finite-size scaling of the spin-glass
susceptibility
\begin{equation}
\label{Eq:ChiScal}
\chi/N  = [\langle q^2 \rangle]_{\rm av} 
\> \sim \> N^{-\Gamma} \> X \left( x \right),
\end{equation}
where $\Gamma$ is an independent critical exponent.  
Reference \cite{decandia:16} adds support to the prediction that above 
$d_{\rm upper}$ one expects
\begin{equation}
\Gamma=\frac{\gamma_{\rm BP}+2\beta_{\rm BP}}{d_{\rm upper}
\nu_{\rm MF}}=\frac{2}{3},
\end{equation}
where BP stands for ``bootstrap percolation'' 
\cite{comment:bp}.  Therefore, a plot of $\chi/ N^{1/3}$ as a function of
$x$ should generate the scaling function $X$ if the correct critical
temperature $T_c$ is used.

Simulations are performed using parallel tempering Monte Carlo
\cite{hukushima:96}.  For each disorder instance we simulate
two independent replicas to compute the spin-glass overlap.  We test
thermalization using the method developed in Ref.~\cite{katzgraber:01}
adapted to the Chimera topology. In all simulations, open boundary
conditions are used to mirror current planar chip designs.

\section{\label{sec:SWno} Chimera lattice with unconstrained small-world
couplers}

The Chimera lattice \cite{bunyk:14} (see Fig.~\ref{Fig:SW:Fig01}) does
not have a finite-temperature spin-glass transition
\cite{katzgraber:14}. As a first step, to verify that a
finite-temperature phase transition in the mean-field universality class
can be induced in a quasiplanar topology, we add unrestricted
small-world couplers to the lattice, depicted in
Fig.~\ref{Fig:SW:Fig02}.  We study lattices with $L\times L$ unit cells
of eight sites each, i.e., $N = 8 \; L^2$ sites.

In the simulations, for each number of sites $N \in \{1152, 1568, 2048,
2592\}$ we thermalize the system for $2^{20}$ Monte Carlo sweeps and
measure over the same number of sweeps. For the parallel-tempering
scheme we use $30$ temperatures in the range $T \in [0.9620,2.3825]$.
Configurational averages are performed over approximately $5000$
samples.

\begin{center}
\begin{figure}[tb]
\vspace{0.05 cm}
\includegraphics[width=\columnwidth]{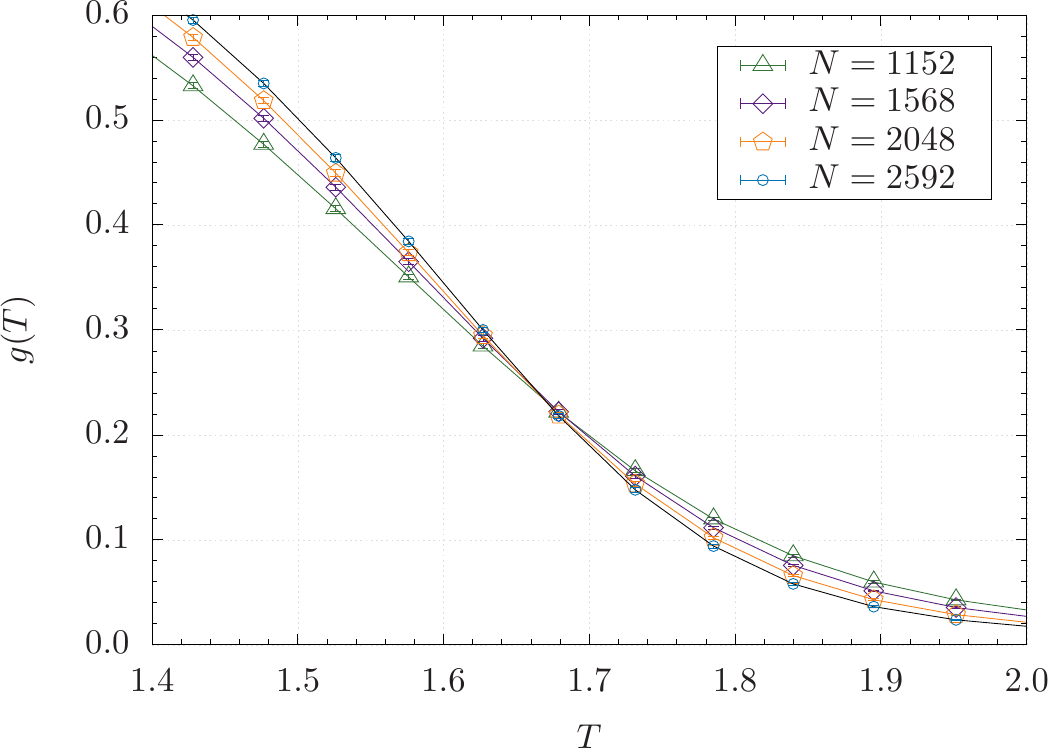}
\caption{
\label{Fig:SW:FigHK01}
Binder cumulant $g$ [Eq.~(\ref{Eq:Binder})] as a function of temperature
$T$ for a spin glass on a Chimera lattice with added unconstrained
small-world interconnects (see  Fig.~\ref{Fig:SW:Fig02}). Data for
different system sizes $N$ cross, suggesting the existence of a
finite-temperature spin-glass transition.
}
\end{figure}
\end{center}

\subsection{Algorithm to add small-world couplers}

The added small-world couplers can take the values $J_{ij} \in \{\pm1\}$
with equal probability. We apply only one restriction to the small-world
couplers; namely, that each site on the native Chimera graph can have at
most {\em one} additional small-world coupler.  The small-world couplers
that connect two qubits that either both lie in the top layer or both
lie in the bottom layer of the Chimera lattice are shown in
Fig.~\ref{Fig:SW:Fig01}. The algorithm to place $n_{\rm SW}/2$ couplers
on the top or bottom layer of Chimera is as follows. Randomly choose two
sites in a given layer that do not yet have a small-world connection
and add a small-world coupler between them until there are $n_{\rm SW}$
couplers either in the top layer or the bottom layer. Note that $n_{\rm SW} \le
N/4$, because the Chimera graph has a two-layer structure and every coupler
connects two sites. In Fig.~\ref{Fig:SW:Fig02} each small-world coupler
is placed at a different level above the top qubit layer (below the
bottom qubit layer) to prevent small-world couplers from
intersecting (a no-crossing constraint). As $N$ increases, the number of
small-world couplers is increased proportionally, so every lattice has
the same fraction of small-world couplers to qubits. This also means
that the number of additional fabrication layers for a chip design grows
accordingly.

\begin{center}
\begin{figure}[tb]
\vspace{0.05 cm}
\includegraphics[width=\columnwidth]{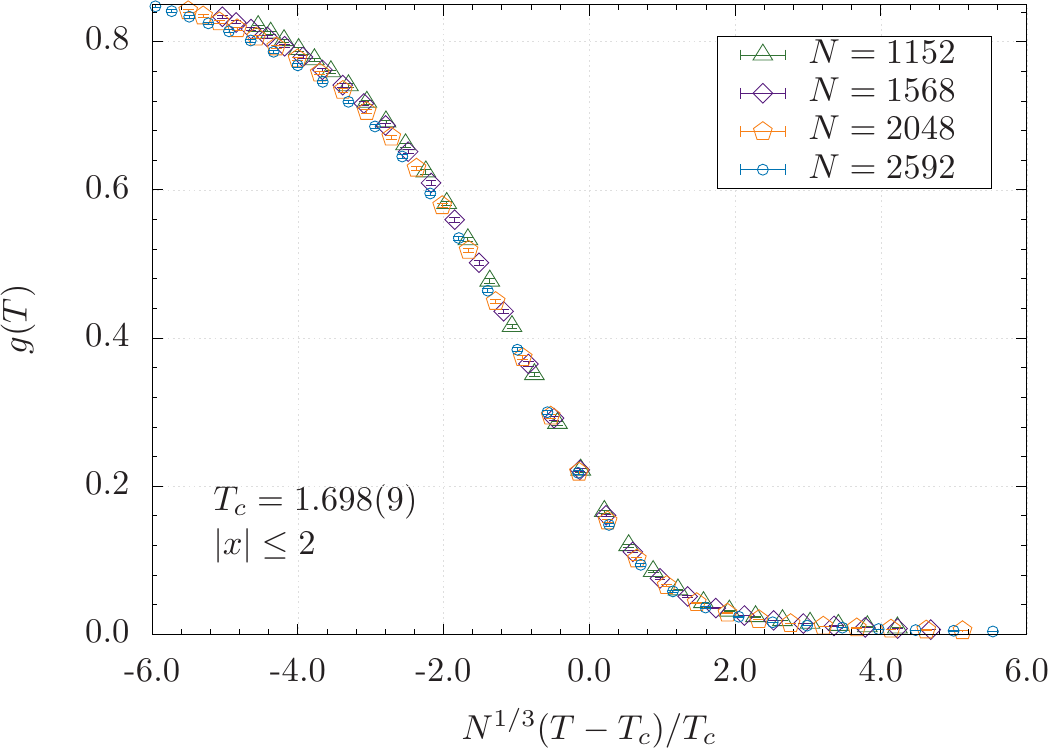}
\caption{
\label{Fig:SW:FigHK02}
Finite-size scaling of the data presented in Fig.~\ref{Fig:SW:FigHK01}
according to Eq.~\eqref{Eq:gScaleff}. The critical exponents are fixed
to the mean-field values and only the critical temperature $T_c$
is a free parameter. The data scale well for $|x| \le 2$ and we estimate
$T_c = 1.698(9)$.
}
\end{figure}
\end{center}

\subsection{Results}

\begin{center}
\begin{figure}[tbh]
\vspace{0.05 cm}
\includegraphics[width=\columnwidth]{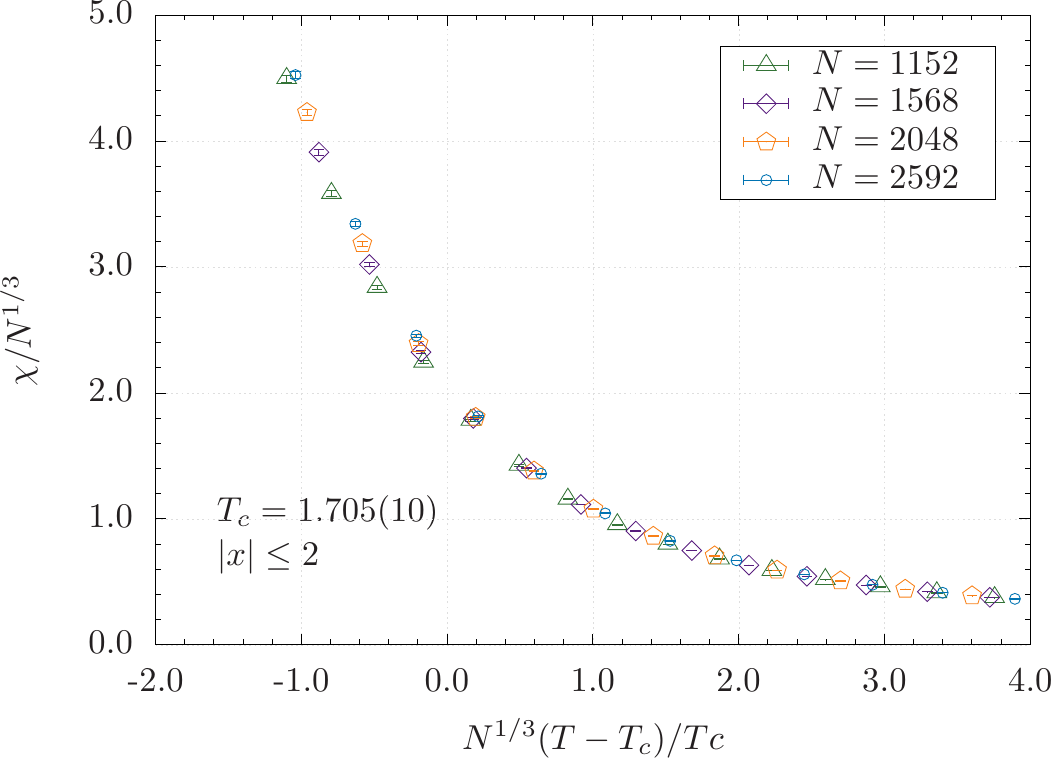}
\caption{
\label{Fig:SW:FigHK03}
Finite-size scaling of the spin-glass susceptibility $\chi$ according to
Eq.~\eqref{Eq:ChiScal}. The critical exponents are fixed to the
mean-field values and only the critical temperature $T_c$ is a
free parameter. The data scale well for $|x| \le 2$ and we estimate
$T_c = 1.705(10)$, in agreement within error bars with the estimate
from the Binder ratio (see Fig.~\ref{Fig:SW:FigHK02}).
}
\end{figure}
\end{center}

Figure \ref{Fig:SW:FigHK01} shows results for the Binder cumulant $g$
for the spin-glass order parameter. The crossings for different $N$
suggest a finite-temperature spin-glass transition. This means that the
addition of small-world couplers to a Chimera lattice generates a
topology where a spin glass can have a finite-temperature transition.
Figure \ref{Fig:SW:FigHK02} shows a finite-size scaling of the data in
Fig.~\ref{Fig:SW:FigHK01}. The scaling of the data is restricted to a
window $|x| \le 2$ and $\nu_{\rm eff} = \nu_{\rm MF} = 3$ fixed. The
data scale well and we estimate $T_c=1.698(9)$.  To corroborate
these results, Fig.~\ref{Fig:SW:FigHK03} shows a finite-size scaling of
the spin-glass susceptibility according to Eq.~\eqref{Eq:ChiScal} with
again the critical exponents fixed to the mean-field values. The data
scale well, and we find $T_c=1.705(10)$, which agrees within error
bars with the estimate obtained from the Binder ratio.  We also perform
the scaling analysis with the different scaling exponents as free
parameters and obtain estimates for the critical temperature that agree
within error bars with the aforementioned estimates, as well as
estimates for the critical exponents that agree with the mean-field
values within error bars.  Therefore, the addition small-world couplers
to a Chimera lattice would allow studies of $J_{ij}$ arrangements that
have characteristics of those of short-range spin glasses with
dimensions $d\ge 6$. This also means that a spin-glass state in a field
might be possible in such architectures.

The data in Figs.~\ref{Fig:SW:FigHK01} -- \ref{Fig:SW:FigHK03} are for
complete Chimera lattices with no vacancies. We also perform simulations
where the Chimera lattices have a small fraction of vacancies---as
commonly found in QAPUs---and averaged over these vacancy arrangements.
Again, the scaled data exhibit a clear finite-temperature spin-glass
transition with exponents expected for a spin glass in the mean-field
regime (not shown).

Figures~\ref{Fig:SW:FigHK01} -- \ref{Fig:SW:FigHK03} are for small-world
connections added to Chimera graphs. These figures should be contrasted
with Fig.~2 in Ref.~\cite{katzgraber:14} for Chimera lattices without
any small-world connections. Figure~2 in Ref.~\cite{katzgraber:14} shows
with use of a finite-size scaling of both the Binder cumulant and the
susceptibility that the transition occurs only at zero temperature.

\section{\label{sec:SWlayer} Chimera lattice with constrained
small-world couplers}

It would be extremely difficult to engineer a QAPU with the
unconstrained small-world interconnects depicted in
Fig.~\ref{Fig:SW:Fig02} because the number of fabrication layers (the
number of layers the small-world couplers with a no-crossing constraint
in every fabrication layer) required would grow rapidly with the number
of qubits $N$.  Therefore, we also investigate the results of placing
further constraints on the couplers added to the underlying Chimera
lattice. In this section we constrain the number of fabrication layers,
with the nonintersecting condition, the added couplers can have.

Fig.~\ref{Fig:SW:Fig06} shows a $K_{4,4}$ Chimera lattice with $L=20$
and added small-world couplers, with four added layers (two above the
layer of the top qubits and two below the layer of the bottom qubits).
The Chimera lattice is depicted as a two-level system. Every
Chimera-lattice unit cell has four qubits in the upper layer and four
qubits in the lower layer, together with each of the top four qubits
connected to every one of the four lower qubits.  The intra-unit-cell
connections in the top layer and the bottom layer are shown. The graph
has free boundary conditions. For details and a better view of the
different couplers and qubits, see Fig.~\ref{Fig:SW:Fig01}. Here we use
the same color coding for the different qubits and couplers as for the
native graph.  In addition, small-world couplers are added in four
additional layers, two above and two below the qubit planes. In this
example, the top qubits are connected by $398$ small-world couplers (red
and black cylinders), and the bottom qubits are connected by 398
small-world couplers (maroon and cyan cylinders).  To keep the
connections from overlapping, every small-world layer is at a different
height above or below the qubit layers.

\subsection{Algorithm to add small-world couplers}
\label{sec:SWlayer2}

\begin{center}
\begin{figure}[tb]
\vspace{0.05 cm}
\includegraphics[width=0.90\columnwidth]{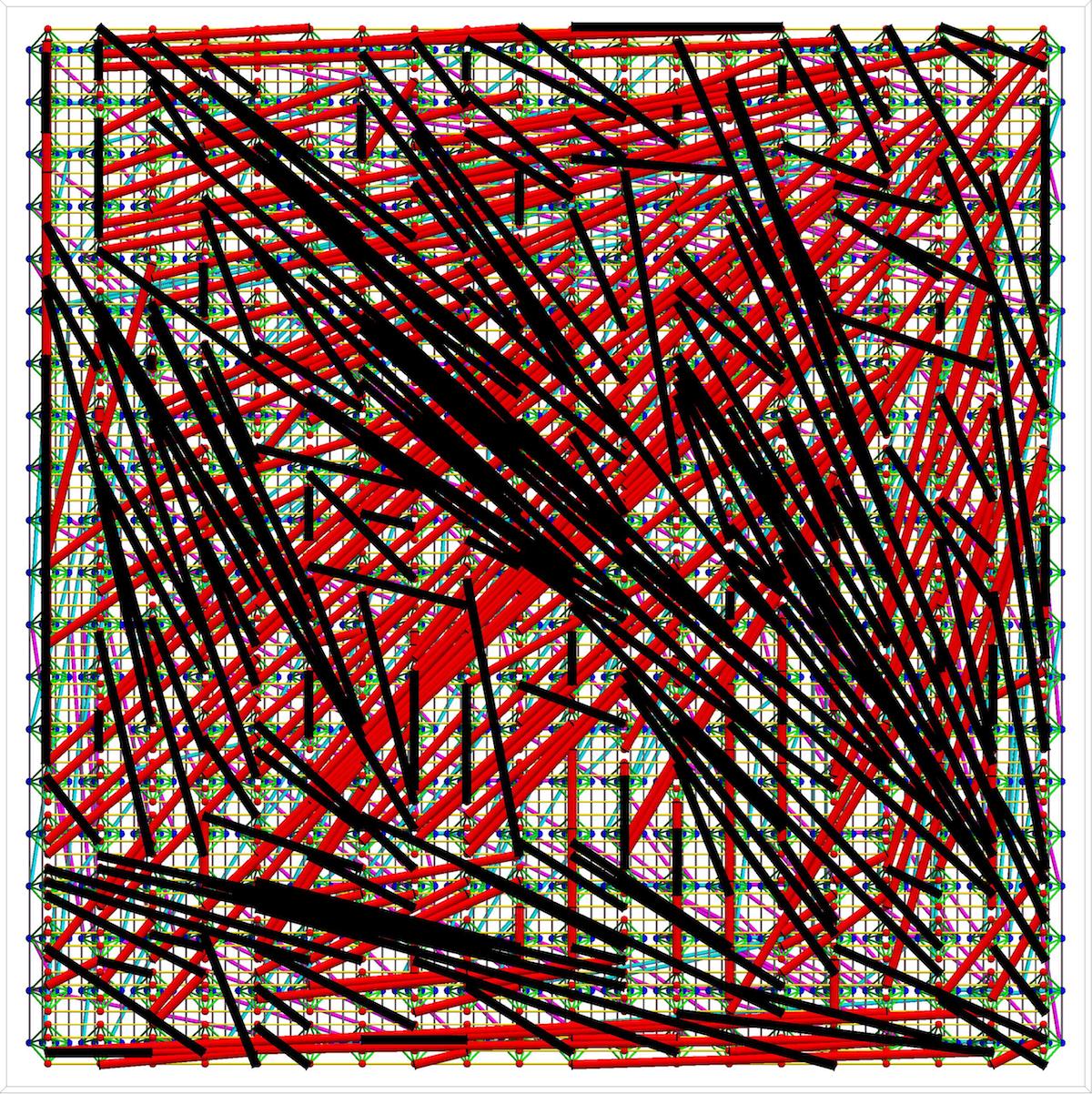}
\caption{
\label{Fig:SW:Fig06}
Chimera lattice ($L = 20$) and added small-world connections constrained
to four layers (two above and two below the standard Chimera topology).
The Supplemental Material contains a three-dimensional interactive
versions of this graph \cite{comment:supp}.
}
\end{figure}
\end{center}

The constraint of the added couplers is that at most one small-world
coupler can be connected to any qubit, that there can be only four
additional layers (two above the top qubit layer and two below the
bottom qubit layer), and the added connections must not intersect within
a layer.  The small-world couplers are added iteratively. For $n_{\rm
SW}/2$ connections for the top- or bottom-layer qubits the iterative
procedure is as follows:
\begin{enumerate} 

\item Select at random two qubits in the layer that do not yet have a
small-world coupler.

\item Attempt to join the two chosen qubits with a small-world coupler.
If the attempted small-world coupler has a slope that is non-negative
and does not intersect any small-world couplers that already have already been
added to the first layer for small-world couplers, add this small-world
coupler to the first small-world layer and to the graph. If the
attempted small-world coupler has a slope that is nonpositive and does
not intersect any other small-world couplers that have already been
added to the second layer for small-world couplers, add this small-world
coupler to the second small-world layer and to the graph. If the
attempted connection cannot be placed in either layer of small-world
couplers without intersecting existing couplers, discard this attempt
and return to step 1.

\end{enumerate} 
The total number of added couplers before no additional ones can be
added depends on the order of the randomly chosen qubits, with the upper
bound $n_{\rm SW}\le N/2$. The slopes of the added couplers are
calculated with the $x$ axis in Fig.~\ref{Fig:SW:Fig06} being
horizontally from left to right, and the $y$ axis being vertically
from top to bottom. An attempted coupler with an infinite slope or zero slope is
allowed to be added to either layer, and an attempt to add it to each
layer is performed.

\begin{center}
\begin{figure}[tb]
\vspace{0.05 cm}
\includegraphics[width=0.90\columnwidth]{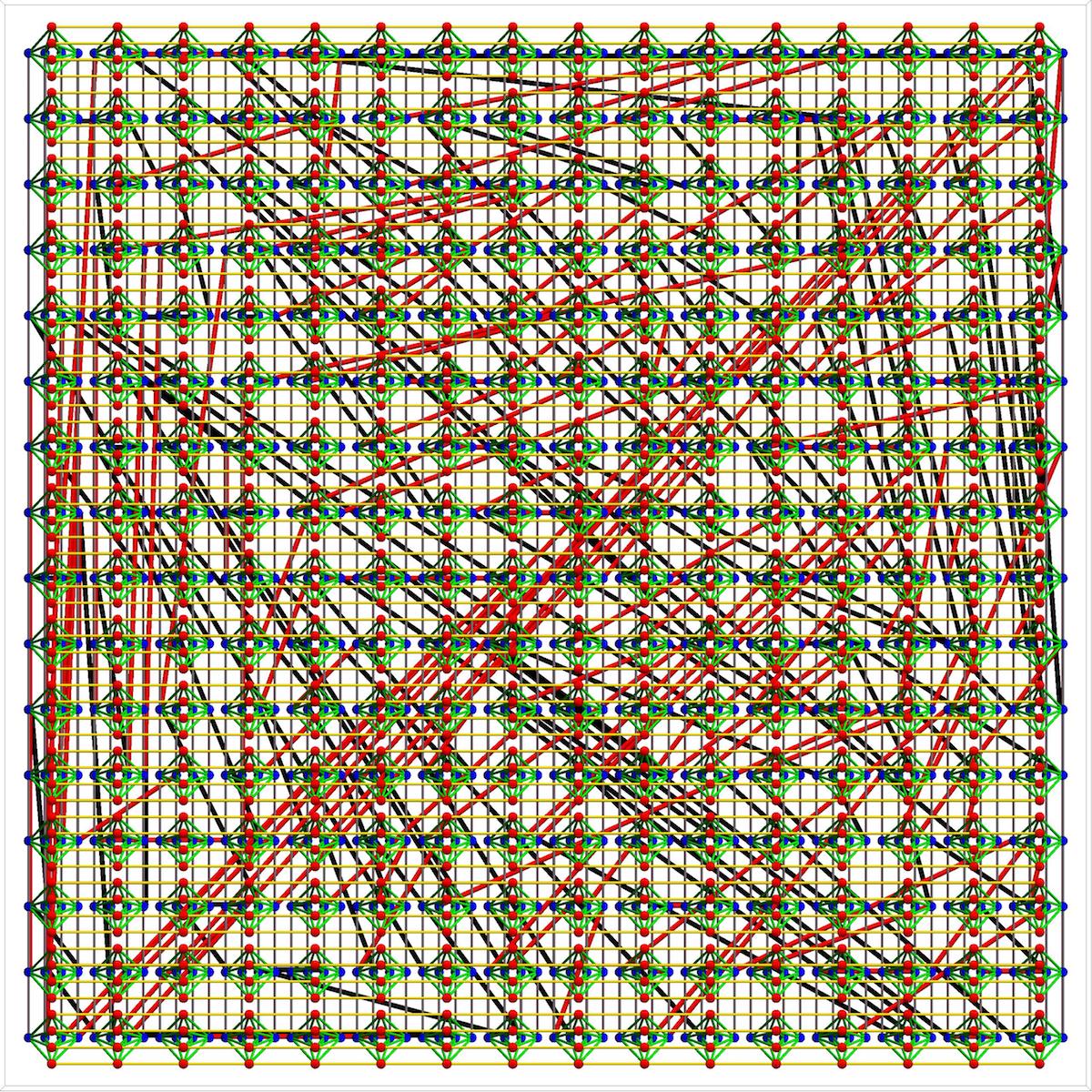}
\caption{
\label{Fig:SW:Fig07}
Chimera lattice ($L = 16$) and added small-world connections constrained
to two layers above the standard qubit planes.  The Supplemental
Material contains a three-dimensional interactive versions of this graph
\cite{comment:supp}.
}
\end{figure}
\end{center}

The fabricated chip of a D-Wave machine with a Chimera graph has a
bottom layer of qubits placed on a substrate surface, with a top layer
of qubits placed in a fabrication layer above the first layer.
Consequently, it may be difficult to fabricate a graph based on a
Chimera lattice with added layers both above and below the fabrication
layers for the Chimera lattice. Therefore, we also perform
calculations to study architectures with added couplers attached only to
the top layer of qubits.  An example of a two-layer system is shown in
Fig.~\ref{Fig:SW:Fig07} for $L=16$. The Chimera graph has the same
structure as in Fig.~\ref{Fig:SW:Fig06}. Small-world couplers are added,
constrained to be in only two layers (both above the Chimera plane of
the top qubits).  The top qubits are connected by $127$ small-world
couplers (red cylinders) in the first layer and $127$ qubits (black
cylinders) in the second layer. To keep the connections from
overlapping, each small-world layer is at a different height above the
top qubit plane.

\subsection{Results}

For the cases of four (Fig.~\ref{Fig:SW:Fig06}) and two
(Fig.~\ref{Fig:SW:Fig07}) additional layers we perform large-scale
Monte Carlo calculations as outlined in Sec.~\ref{sec:SWno}.  Averaging
over both coupler values and different coupler distributions results
again in a finite-temperature spin-glass transition. However, the latter
is unrealistic because the chip design should be static. Fixing the
coupler positions and performing simulations where the configurational
average is over only coupler values results in some cases where a
possible transition is observed and others where this is not the case
(not shown).  Clearly, the strong fluctuations in the position of the
couplers thus results in lattices that might have been at times
mean-field-like and at other times not. This effect is far-more
pronounced in the system with two additional layers, which is why we
next focus on the system with four additional layers. Furthermore,
additional constraints on the small-world couplers are clearly needed to
ensure that a reproducible spin-glass transition is found and to reduce
these strong corrections to scaling.

\section{\label{sec:SWlayerangle} Chimera lattice with angle-constrained
small-world couplers}

\begin{center}
\begin{figure}[tb]
\vspace{0.05 cm}
\includegraphics[width=0.90\columnwidth]{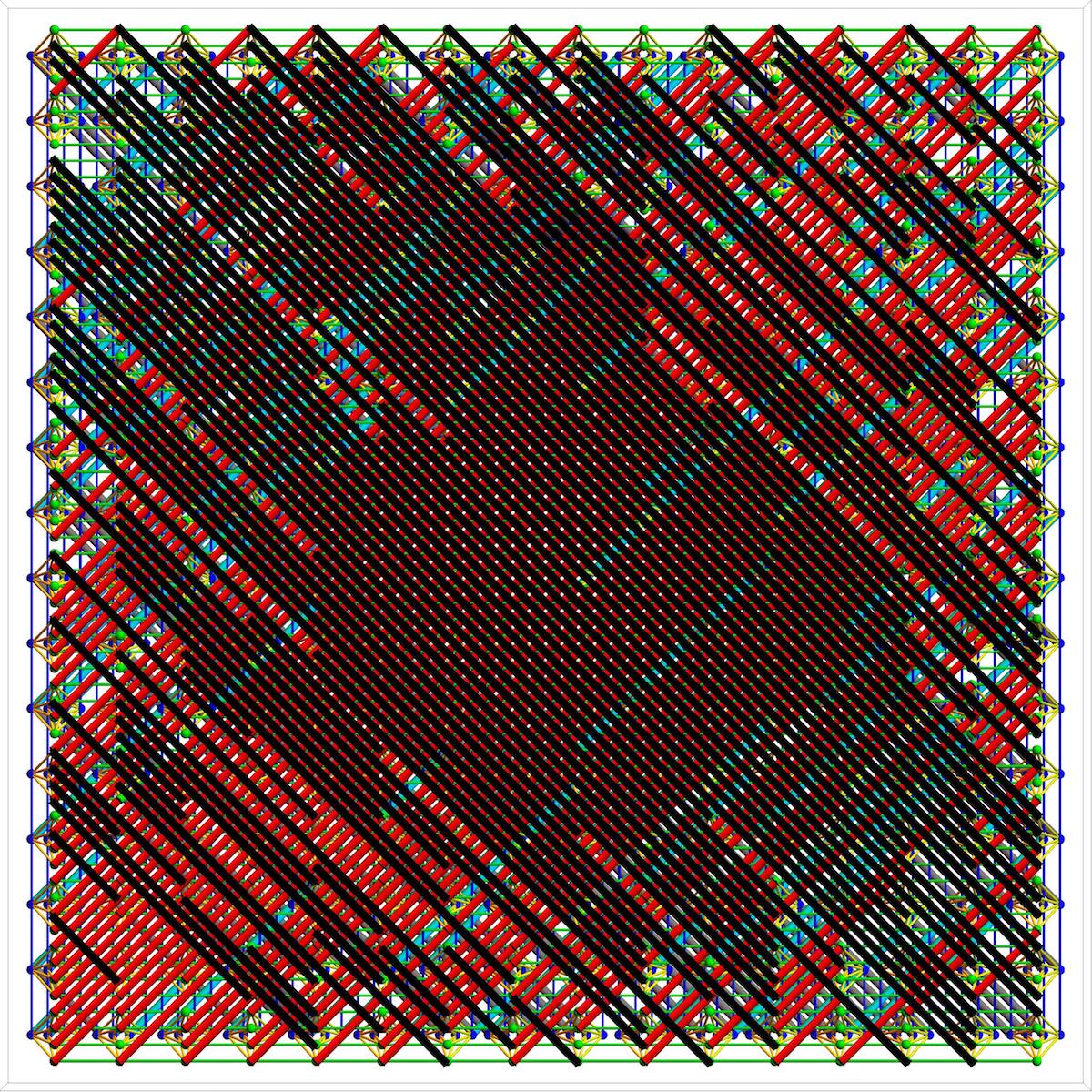}
\caption{
\label{Fig:SW:Fig08}
Chimera lattice ($L=16$) with added small-world couplers on four layers
(two on top and two below), with the connections constrained to diagonal
directions.  The lattice has $N = 2048$ qubits, $12 032$ couplers, and
additional $n_{\rm SW} = 1020$ small-world couplers ($255$ per layer).
The Supplemental Material contains a three-dimensional interactive
version of this graphs \cite{comment:supp}.
}
\end{figure}
\end{center}

Fabricating systems as depicted in Fig.~\ref{Fig:SW:Fig06} or
Fig.~\ref{Fig:SW:Fig07} may still be difficult. Furthermore, tuning the
qubits to be within the desired specifications for a QAPU may be
extremely difficult for such graphs. In an effort to also reduce the
fluctuations and seemingly random results obtained in
Sec.~\ref{sec:SWlayer}, we study the effects of the further constraint
of all added couplers having slopes of either $+1$ or $-1$ (angles
formed of $+\pi/4$ or $-\pi/4$).  The slopes are calculated only from
the unit cells, not from specific positions of the qubit locations
within a unit cell. An example graph with four added layers is shown in
Fig.~\ref{Fig:SW:Fig08}. The procedure to add the small-world couplers
is the same as outlined in Sec.~\ref{sec:SWlayer2}, but with the added
rejection step the first-layer added connections must form an angle of
$+\pi/4$ and the second-layer added connections must have an angle of
$-\pi/4$.

We perform large-scale Monte Carlo simulations for system sizes $N = 8
L^2$ with $L \in \{16, 18, \ldots, 28, 30\}$ (i.e., $N = 2048, \ldots,
7200$) and thermalize the system for $2^{23}$ Monte Carlo sweeps and measure for
the same number of sweeps.  For the parallel-tempering scheme we use
$30$ temperatures in the range $T \in [1.5,2.3]$.  Configurational
averages are performed over approximately $5000$ samples for each system
size. The additional number of small-world couplers for each system size
is $n_{\rm SW} = 4(L^2 - 1)$.

\begin{center}
\begin{figure}[tbh]
\vspace{0.05 cm}
\includegraphics[width=\columnwidth]{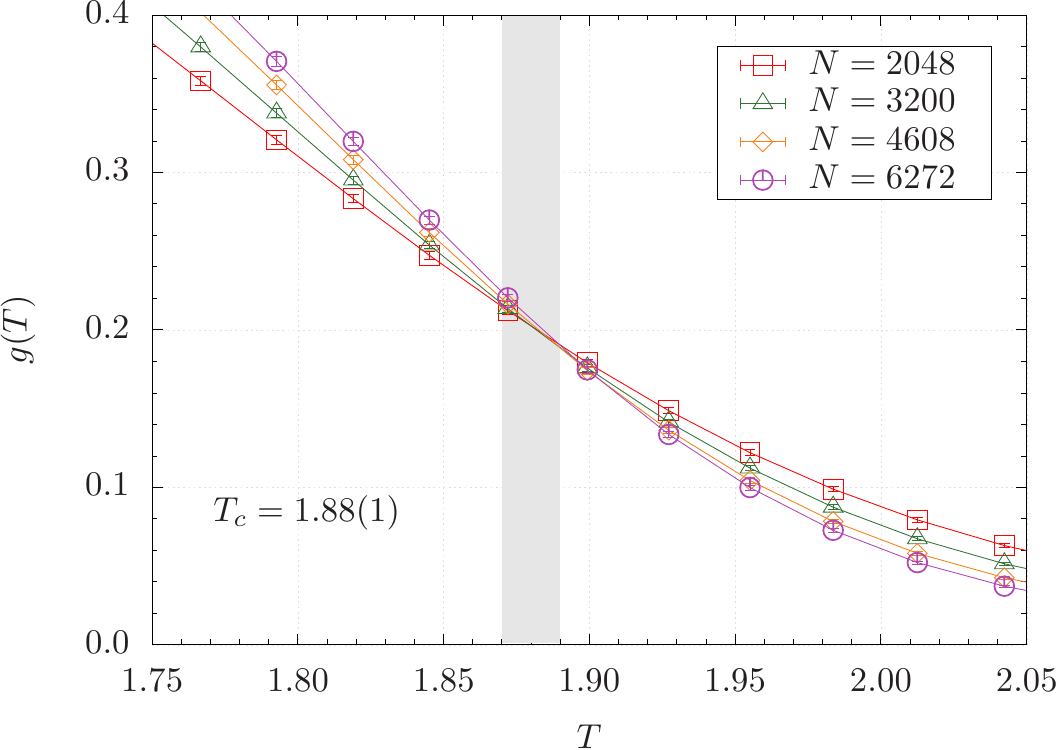}
\caption{
\label{Fig:SW:Fig09}
Binder cumulant $g$ [Eq.~(\ref{Eq:Binder})] as a function of temperature
$T$ for a spin glass on a Chimera lattice with added angle-constrained
small-world interconnects (see Fig.~\ref{Fig:SW:Fig08}). Data for
different system sizes $N$ cross, suggesting the existence of a
finite-temperature spin-glass transition.  Note that corrections to
scaling are large.
}
\end{figure}
\end{center}

\begin{center}
\begin{figure}[tbh]
\vspace{0.05 cm}
\includegraphics[width=\columnwidth]{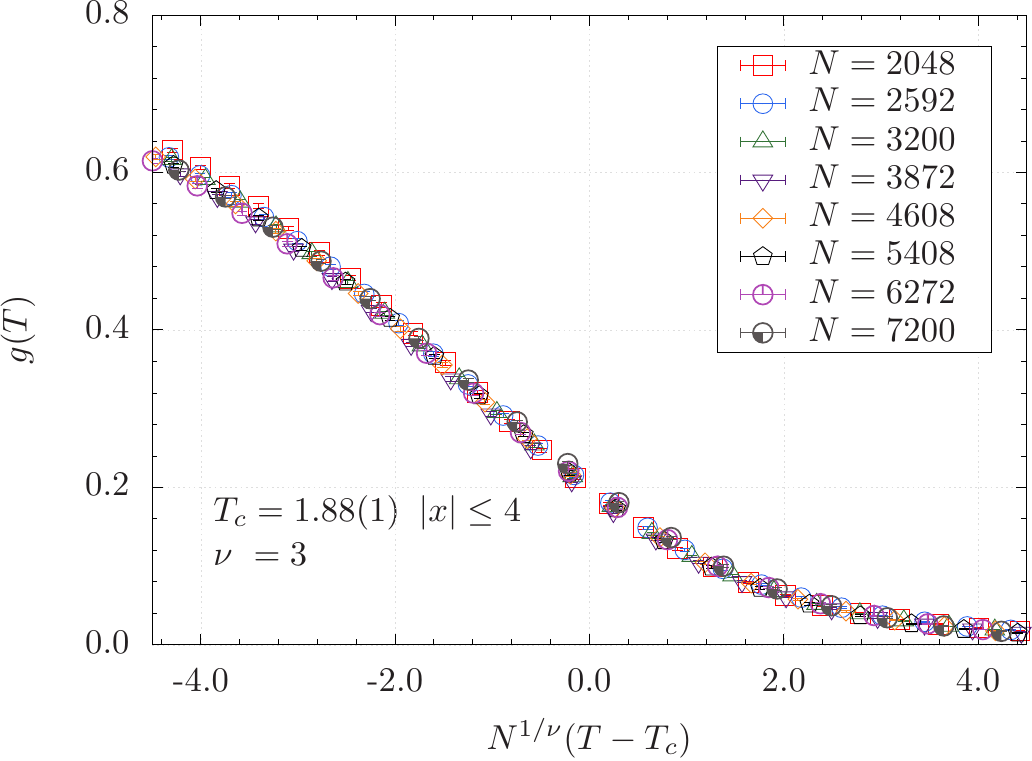}
\caption{
\label{Fig:SW:Fig10}
Finite-size scaling of the data presented in Fig.~\ref{Fig:SW:Fig09}
according to Eq.~\eqref{Eq:gScaleff}. The critical exponents are fixed
to the mean-field values and only the critical temperature $T_c$
is a free parameter. The data scale well for $|x| \le 4$, and we estimate
$T_c = 1.88(1)$.
}
\end{figure}
\end{center}

\begin{center}
\begin{figure}[tbh]
\vspace{0.05 cm}
\includegraphics[width=\columnwidth]{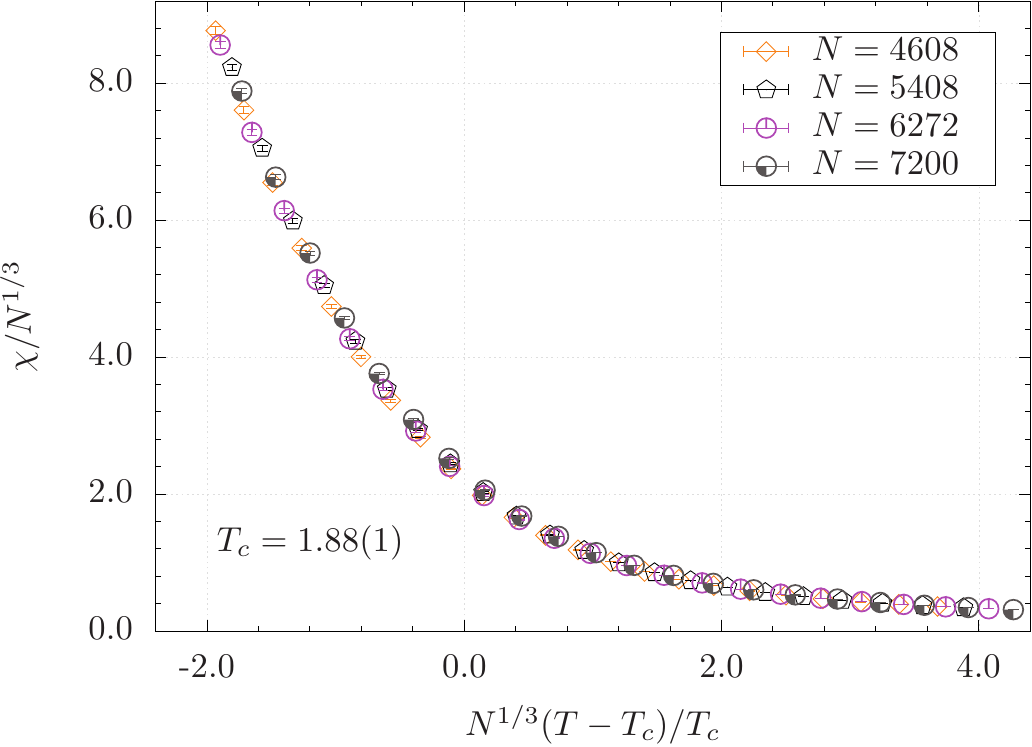}
\caption{
\label{Fig:SW:Fig11}
Finite-size scaling of the spin-glass susceptibility $\chi$ according to
Eq.~\eqref{Eq:ChiScal}. The critical exponents are fixed to the
mean-field values and only the critical temperature $T_c$ is a
free parameter. The data scale well for $|x| \le 4$, and we estimate
$T_c = 1.88(1)$, in agreement within error bars with the estimate
from the Binder ratio (see Fig.~\ref{Fig:SW:Fig09}).
}
\end{figure}
\end{center}

Figure \ref{Fig:SW:Fig09} shows results for the Binder cumulant $g$ for
the spin-glass order parameter. The crossings for different $N$ suggest a
finite-temperature spin-glass transition for $T \approx 1.88$.  Compared
with the unrestricted case, corrections to scaling are evident; however,
they are much smaller than in Sec.~\ref{sec:SWlayer}. This is
likely also enhanced due to the open boundaries. We decide not to use
periodic boundaries as we want to demonstrate that the proposed
approach works for realistic hardware graphs.  This means that even for
highly restricted small-world couplers a finite-temperature spin-glass
transition can be induced for a spin glass on the Chimera lattice.  To
overcome the strong corrections to scaling, we fit the data in
Fig.~\ref{Fig:SW:Fig09} with the known mean-field critical exponents
(namely, $\nu_{\rm eff}=3$ and $\Gamma= 2/3$) and focus only on the largest
system sizes simulated.  Figure \ref{Fig:SW:Fig10} shows a finite-size
scaling of the data for the Binder ratio. The data have been scaled for
$|x| \le 4$, and scale well. We also perform a scaling of the data
with the critical exponent $\nu$ as a variable. The results obtained
agree with the exact mean-field value within error bars; however, these
are relatively large. Figure \ref{Fig:SW:Fig11} shows scaled data for
the susceptibility $\chi$ obtained with the expected mean-field exponents.  The
scaling is good, and from the scaling of both the Binder ratio and the
susceptibility we estimate $T_c = 1.88(1)$. Again, allowing the
exponents to fluctuate results in estimates with large error bars.

\section{\label{sec:SWsquare} Square lattices with angle-constrained
small-world couplers}

\begin{center}
\begin{figure}[tb]
\vspace{0.05 cm}
\includegraphics[width=0.90\columnwidth]{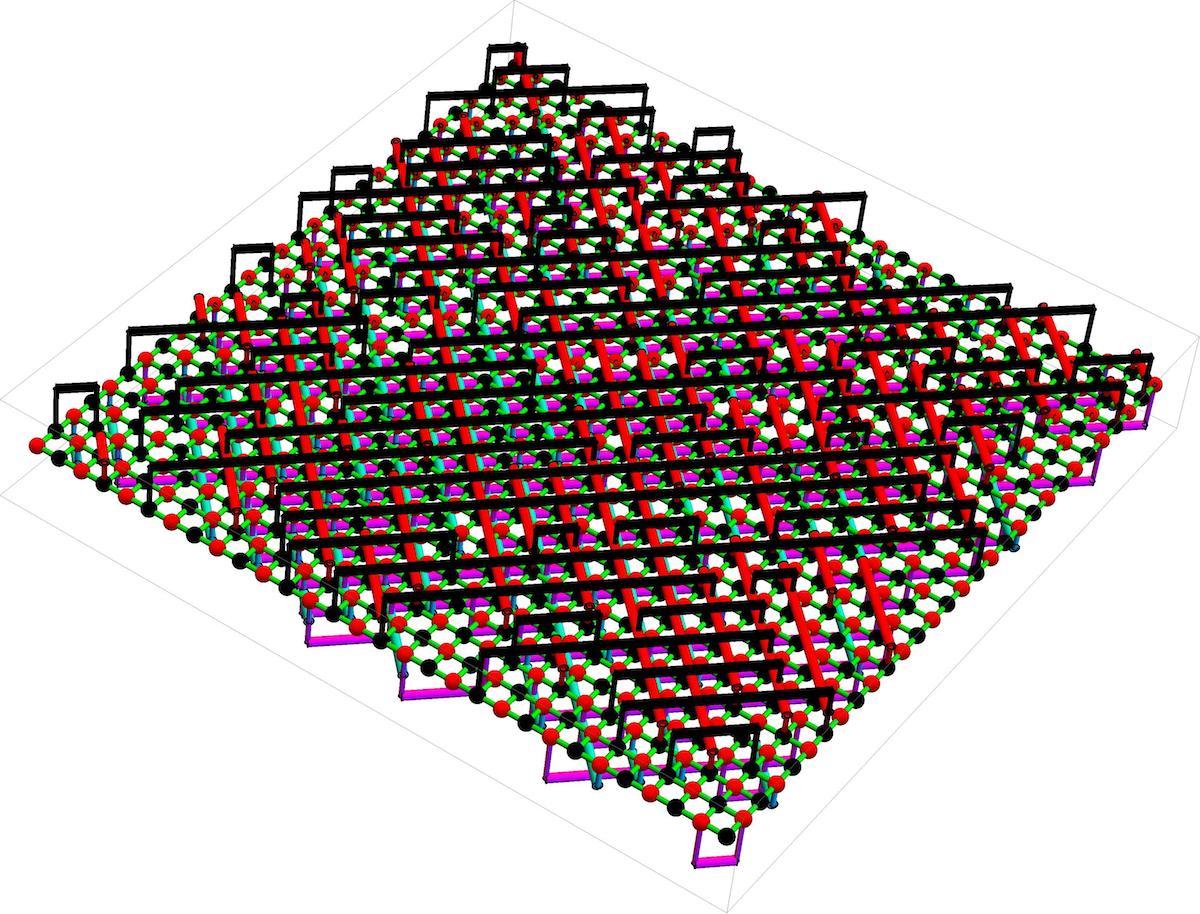}
\caption{
\label{Fig:SW:Fig12}
Square lattice of linear size $L=30$ with free boundary conditions with
added angle-constrained small-world couplers.  The lattice is bipartite,
with the two sublattices having qubits denoted by red or black spheres.
The square-lattice couplings between qubits are denoted by green bonds.
The small-world couplers are constrained to be along only the diagonal
and are composed of four layers, each at a different height and of a
different color (red, black, cyan, magenta), and each with $50$
small-world couplers.  The Supplemental Material contains a
three-dimensional interactive version of this graph \cite{comment:supp}.
}
\end{figure}
\end{center}

We demonstrate the generality of the approach and study the effects of
open boundary conditions and constrained additional couplers on the
scaling and perform large-scale Monte Carlo simulations on square
lattices. As for the Chimera lattice, we use free boundary conditions.
The lattices have $N=L^2$ qubits, with $L$ the linear size. Each qubit
that is not on the lattice boundary has four couplers to its
nearest-neighbor qubits. The added small-world couplers are constrained
to four layers (two on each side of the lattice) and to having slopes of
$\pm 1$ (i.e., along the diagonals as in Sec.~\ref{sec:SWlayerangle} for
Chimera lattices).  The bipartite (red-black-checkerboard) nature of the
square lattice is used to allow for the added connections between only
red-red qubits or black-black qubits. This mimics the top-bottom
structure used in attaching additional couplers to the qubits of the
Chimera lattices.  An example square lattice with added small-world
couplers is shown in Fig.~\ref{Fig:SW:Fig12}. For all system sizes $N$
simulated, the data are thermalized for $2^{18}$ Monte Carlo sweeps and
averaged over an additional $2^{18}$ sweeps. In the parallel-tempering
method we use $30$ temperatures in the range $T \in [0.5,1.3]$.
Configurational averages are performed over approximately $5000$ random
coupler settings with the location of the small-world couplers fixed for
each system size. For a lattice of size $N = L^2$, an additional $n_{\rm
SW} = N/4$ small-world couplers are added, with $n_{\rm SW}/4$ per
additional small-world layer.

\begin{center}
\begin{figure}[tb]
\vspace{0.05 cm}
\includegraphics[width=\columnwidth]{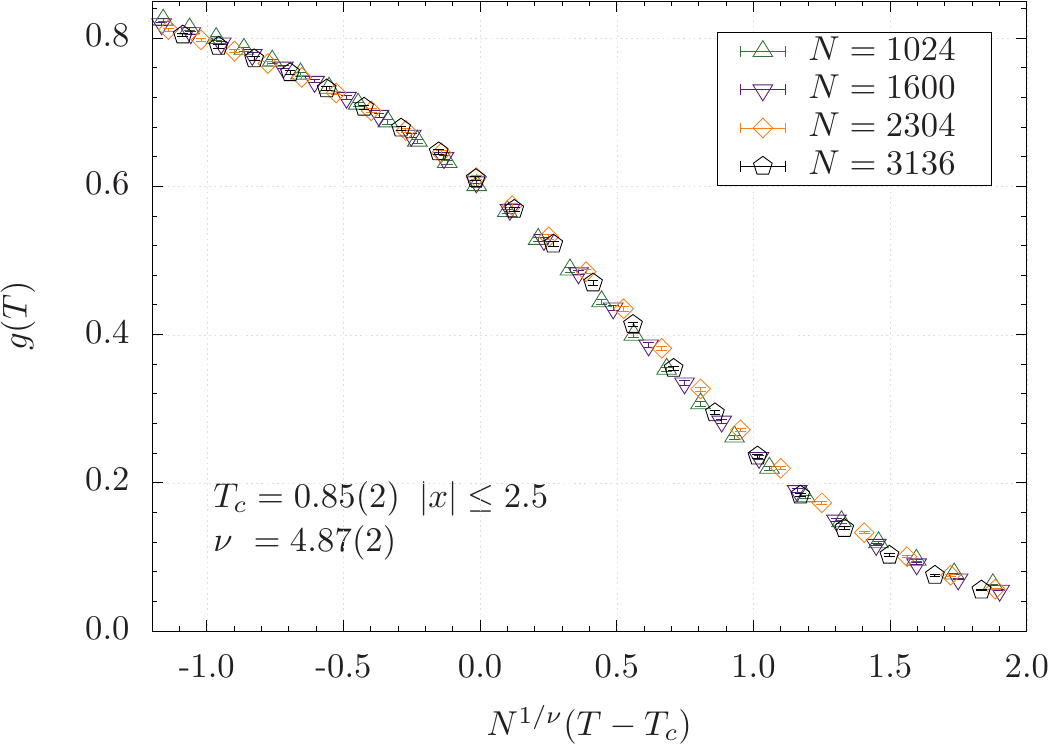}
\caption{
\label{Fig:SW:Fig13}
Finite-size scaling of the Binder cumulant $g$ according to
Eq.~\eqref{Eq:gScaleff} for a square lattice with angle-constrained
small-world couplers. The critical exponents are allowed to fluctuate.
A scaling for $|x| \le 2.5$ results in $T_c =
0.85(2)$ with $\nu = 4.87(2)$.
}
\end{figure}
\end{center}

Figure \ref{Fig:SW:Fig13} shows a finite-size scaling of the spin-glass
Binder ratio $g$ when both $T_c$ and $\nu_{\rm eff}$ are allowed
to be free parameters. The open boundary conditions, together with the
constrained added bonds, lead to large corrections to scaling. The data
only scale for $\nu_{\rm eff} = 4.87(2)$---a value far from the known
mean-field value. A scaling analysis of the data for the susceptibility
$\chi$ was not possible. Corrections to scaling for the square lattice
are much greater than for the Chimera lattice when open boundary
conditions are used, likely because the number of variables is $8$ times
smaller.  Nevertheless, the data show there is a finite spin-glass
transition temperature.

\section{\label{sec:DisCon} Discussion and Conclusions}

We investigate how small-world interactions can lead to improved
quantum annealer designs.  In particular, we perform large-scale
Monte Carlo simulations of spin-glass models with added small-world
couplers. The data are analyzed with use of a finite-size scaling for both the
Binder cumulant $g$ and the spin-glass susceptibility $\chi$. The
scaling allows us to determine whether or not there is a
finite-temperature spin-glass transition, together with the associated
critical exponents, which can be used to determine the ``mean-field-ness''
of the Ising spin-glass model on the underlying graph.

Spin-glass models on graphs with added small-world connections are 
shown to have a finite spin-glass transition temperature, and the
associated critical exponents are those of a mean-field spin glass (or a
spin glass above the upper critical dimension $d_{\rm upper}=6$).  Thus,
small-world graphs should enable easier embedding of harder
spin-glass problems. The ability to more-easily embed harder problems,
in turn, should aid in the quest for adiabatic quantum machines to show
quantum speedup.  Furthermore, we investigate how the addition of
a few extra layers in a chip-fabrication process, necessitating
enforcing constraints on the small-world connections, may also lead to
spin-glass critical behavior similar to that of unconstrained
small-world connections.

We first investigate a $K_{4,4}$ Chimera lattice with added
small-world couplers. It is known that a spin glass on a $K_{4,4}$
Chimera lattice---such as in Fig.~\ref{Fig:SW:Fig01}---does not exhibit
a finite-temperature spin-glass transition \cite{katzgraber:14}.  In
Sec.~\ref{sec:SWno} we demonstrate that adding of small-world couplers to
the Chimera lattice, as in Fig.~\ref{Fig:SW:Fig02}, does result in a
finite spin-glass transition temperature.  Furthermore, the critical
exponents obtained via a finite-size scaling analysis are those expected
for an Ising spin glass with a dimension $d\ge d_{\rm upper}=6$.
Therefore, adding small-world couplers to a Chimera lattice should
enable the study of more complex optimization problems on a quantum
annealer.

To bring the graphs into the realm where engineering is
possible, we examine constraining the small-world couplers.  For the
underlying Chimera lattice, in Sec.~\ref{sec:SWlayer} the added
connections are constrained to either four or two additional layers,
resulting in disorder-dependent results---an undesirable outcome.  In
Sec.~\ref{sec:SWlayerangle} the small-world couplers are further
constrained to be parallel to the diagonals in the lattice.  In this
case we obtain again a finite transition temperature in the mean-field
regime. We also study the effect of such constrained bonds on an
underlying square lattice in Sec.~\ref{sec:SWsquare} (Fig.~\ref{Fig:SW:Fig12}), where we observe a finite-temperature
spin-glass transition. However, the corrections to scaling are large and we
are unable to verify that the exponents are mean-field-like.

One can estimate the critical temperature of the spin glass with
small-world couplers and compare this with the refrigerator temperature of
current quantum annealing machines.  For the current D-Wave machine
\cite{dwaveWEB}, the D-Wave 2000Q with approximately 2000 qubits, the
refrigerator temperature is $T_{\rm fridge}=0.015$K, while the
the average spin-glass coupling $\left\vert J_{ij}\right\vert$ is just
above $10$GHz (i.e., $0.5$K). The anticipated spin-glass temperature due
to the added small-world couplers is approximately the same order of
magnitude as $\left\vert J_{ij}\right\vert$.  Hence, the existence of the
finite-temperature spin-glass transition due to small-world connections
should be important even for today's early QAPUs, and should contrast
sharply with having a graph such as the D-Wave Chimera lattice of the
2000Q device, which does not have a finite-temperature spin-glass
transition $0=T_{c}<T_{\rm fridge}$.  Therefore, for future QAPUs one
expects the addition of small-world connections will lead to a
relationship between these temperatures of $0<T_{\rm fridge}<<T_{c}$.

In this study we only analyze $2$-local Hamiltonians.  Future annealing
machines may have higher-body couplers, and hence the architecture would
be associated with hypergraphs. Even in this case, if there is an
underlying unit cell with a fixed number $n_{\rm cell}$ of qubits and
the hypergraph has a total number of sites $N=n_{\rm cell}L^d$ (based on
a $d$-dimensional lattice) we anticipate that adding small-world
couplers could result in a finite-temperature spin-glass transition. For
example, in the case of a $3$-local lattice a planar triangular lattice
has a zero-temperature spin-glass transition \cite{katzgraber:09c}.  The
addition of small-world couplers would likely result in a
finite-temperature transition into a spin-glass state.

Further analysis could investigate how the recently introduced idemetric
property (wherein most distances between nodes are almost the same)
relates to QAPU designs that can be engineered in a reasonable fashion
\cite{barmpalias:18x}.

Another advantage of lattices enhanced with small-world couplers and
their mean-field-like universality class is the possible existence of a
spin-glass state in a field. This could be desirable for the generation
of more complex synthetic native benchmark problems with local biases
and might even help elucidate the behavior of spin glasses in a field
via quantum annealing.

\begin{acknowledgments}

H.G.K.~thanks Andrew Ochoa and Zheng Zhu for multiple discussions and
acknowledges the ARC Centre for Excellence in All-Sky Astrophysics in 3D
(ASTRO 3D) East Coast Writing Retreat 2018 for support in preparing the
manuscript. M.A.N.~would like to thank Hans De~Raedt, Fenping Jin,
Yaroslav Koshka, Kristel Michielsen, and Dilina Perera, for useful
discussions.  H.G.K.~acknowledges support from the NSF (Grant
No.~DMR-1151387), and thanks Aji Panca and Wilbur Scoville for
inspiration. M.A.N.~thanks the Faculty of Mathematics and Physics at
Charles University for its hospitality during a sabbatical stay, and
funding as a Fulbright Distinguished Chair from the Czech Fulbright
Commission. M.A.N.~carried out part of this work at the Aspen Center for
Physics, which is supported by National Science Foundation (Grant
No.~PHY-1607611).  We thank the Texas Advanced Computing Center
at the University of Texas at Austin for providing high-performance computing
resources
(Stampede cluster) and Texas A\&M University for access to its Ada,
and Lonestar clusters.  H.G.K.'s research is based upon work supported
by the Office of the Director of National Intelligence (ODNI),
Intelligence Advanced Research Projects Activity (IARPA), via
Interagency Umbrella Agreement No.  IA1-1198. M.A.N.'s research is
based upon work supported by the Air Force Research Laboratory (AFRL)
under agreement number FA8750-18-1-0096.  The views and conclusions
contained herein are those of the authors and should not be interpreted
as necessarily representing the official policies or endorsements,
either expressed or implied, of the ODNI, IARPA, AFRL, or the
U.S.~Government. The U.S.~Government is authorized to reproduce and
distribute reprints for Governmental purposes notwithstanding any
copyright annotation thereon.

\end{acknowledgments}

\bibliography{refs,comments}

\end{document}